# Effective Personalized Recommendation in Collaborative Tagging Systems

Zi-Ke Zhang
Department of Physics
University of Fribourg, Switzerland
zhangzike@gmail.com

Tao Zhou
Lab of Infophysics
University of Science and Technology of China
zhutou@ustc.edu

## ABSTRACT
Recently, collaborative tagging systems have attracted more and more attention and have been widely applied in web systems. Tags provide highly abstracted information about personal preferences and item content, and are therefore potential to help in improving better personalized recommendations. In this paper, we propose a tag-based recommendation algorithm considering the personal vocabulary and evaluate it in a real-world dataset: *Del.icio.us*. Experimental results demonstrate that the usage of tag information can significantly improve the accuracy of personalized recommendations.

## 1. INTRODUCTION
The exponential growth of web information has brought us into an information overload era: We face too much data and sources to be able to find out those most relevant and interesting for us. Evaluating all these alternatives by ourselves is not possible. As a consequence, an urgent problem is how to automatically find out the relevant items for us. Internet search engine [3] provides us a useful tool to find out those information and it achieves great success over the last decade. However, it does not take into account personalized information and returns the same results for people with far different habits. Comparatively, *recommender system* [15], adopting knowledge discovery techniques to provide personalized recommendations, is now considered to be the most promising way to efficiently gather the useful information. Thus far, recommender systems have successfully found applications in e-commerce [16], such as book recommendations in *Amazon.com* [11], movie recommendations in *Netflix.com* [2], video recommendations in *TiVo.com* [1], and so on.

One of the most prominent techniques of recommender systems is Collaborative Filtering (CF), where a user is recommended items that people with similar tastes and preferences liked in the past. Despite its success, the performance of CF is strongly limited by the sparsity data. Thus, a number of researches devoted to integrate additional information, such as user profiles [10], item content [14] and attributes [21], to filter out possibly irrelevant recommendations. However, these applications are usually strongly restricted to respect personal privacy, or limited due to the lack of available content information.

Collaborative tagging systems (CTSes), allowing users to freely assign tags to their collections, provide promising possibility to better address the above issues. CTSes require no specific skills for user participating, thus can overcome the limitation of vocabulary domains and size, widen the semantic relations among items and eventually facilitate the emergence of *folksonomy* [9]. In addition, tags can be treated as abstracted content of items. Especially, tags are given by users themselves and thus in somehow represent the personal vocabulary and preferences. In this paper, we propose a tag-based recommendation algorithm to that takes into account the personal vocabulary. We use one benchmark data set, *Del.icio.us*, to evaluate our algorithm. Experimental results demonstrate that the usage of tag information can significantly improve the accuracy of recommendations.

The rest of this paper is organized as follows. Section 2 reviews the related work. In Section 3 we introduce our proposed algorithm and report the experimental results. Finally, we summarize this paper and outline some open issues for future research in Section 4.

## 2. RELATED WORK
Recently, many efforts have been addressed in understanding the structure, evolution [4] and usage patterns [7] of CTSes. A considerable number of algorithms are designed to recommend tags to users, which may be helpful for better organizing, discovering and retrieving items [9, 12, 20].

Table 1: Basic information of the data set.

| Value | Description |
|---|---|
| 9,991 | number of users |
| 243,737 | number of items |
| 102,732 | number of tags |
| 1,257,908 | number of user-item relations |
| 4,391,073 | accumulative number of tags |

The current work focuses on a relevant yet different application of CTSes, that is, to provide personalized item recommendations with the help of tag information. Schenkel *et al.* [17] proposed an incremental threshold algorithm taking into account both the social ties among users and semantic relatedness of different tags, which performs remarkably better than the algorithm without tag expansion. Nakamoto *et al.* [13] created a tag-based contextual collaborative filtering model, where the tag information is treated as the users' profiles. Tso-Sutter *et al.* [22] proposed a generic method that allows tags to be incorporated to the standard collaborative filtering, via reducing the ternary correlations to three binary correlations and then applying a fusion method to re-associate these correlations. Chi *et al.* [5] presented a model considering probabilistic polyadic factorization for personalized recommendation. Shepitsen *et al.* [19] proposed a tag clustering-based method to improve the algorithmic accuracy. Zhang *et al.* [24] presented a diffusion-based hybrid algorithm for personalized recommendation in CTSes. Shang *et al.* [18] proposed a hybrid collaborative filtering algorithm on user-item-tag tripartite graphs.

## 3. ALGORITHM AND EXPERIMENTS

In this paper, we adopt a weighted variant of diffusion-based method proposed in [25], where the weights are given according to personal vocabulary in CTSes. A CTS consists of three sets, for users $U = \{U_1, U_2, \cdots, U_n\}$, items $I = \{I_1, I_2, \cdots, I_m\}$, and tags $T = \{T_1, T_2, \cdots, T_s\}$, respectively. Actually, it is easy to understand that different users may consider differently for the same item, and such difference can be characterized to some extent by looking into the different usage pattens of tags. Although those tags are freely given, people are supposed to give their most favorite words to describe their best collections. A latent assumption is that the more frequently a user uses a tag, the more likely the user likes this tag as well as the items labeled with it. On the other hand, users are not willing to give too many tags for a single item.

### 3.1 Algorithm

In this subsection, we introduce a simple way that utilizes the tag information to provide better recommendations. As mentioned above, we will consider two factors: (i) the frequency of each tag used by each user; (ii) the number of tags assigned with a single item. Since our aim is to find the most relevant items for a particular user, so-called personalized recommendation, we will describe our algorithm for a target user $U_i$. The algorithm can be expressed in following steps:

**Step 1:** Define the initial value vector $\vec{f}$ for all the items, whose element reads:

$$f_j = \frac{1}{\sum_{j=1}^{m} \sum_{s'=1}^{|T_{ij}|} K(t_{is'})} \sum_{s=1}^{|T_{ij}|} K(t_{is}), \quad (1)$$

where $|T_{ij}|$ denotes the number of tags that $U_i$ has assigned to item $I_j$, and $K(t_{is})$ is the number of times tag $t_s$ has been used by $U_i$.

**Step 2:** Distribute the value of each item evenly to the users who collect it, then the value a user $U_l$ will receive reads:

$$r_l = \sum_{j \in \Gamma(U_l)} \frac{f_j}{d(I_j)}, \quad (2)$$

where $\Gamma(U_l)$ denotes the set of items collected by $U_l$, and $d(I_j)$ is the degree of $I_j$ in the user-item bipartite graph.

**Step 3:** Redistribute the value of each user $U_i$ to his/her collections according to the weight defined in Step 1. Then the final value vector $\vec{f}'$ of items will be summarized as:

$$f'_j = \sum_{k=1}^{|U_{I_j}|} \frac{r_k}{\sum_{j=1}^{m} \sum_{s'=1}^{|T_{kj}|} K(t_{ks'})} \sum_{s=1}^{|T_{kj}|} K(t_{ks}),, \quad (3)$$

where $|U_{I_j}|$ is the number of users collected item $I_j$.

The above procedure constitutes of a *mutual reinforcement* process that allows the values transferred between users and items. At the first step, we highlight the items selected by $U_i$ and assign each of them with an initial value according to $U_i$'s tagging activities. Step 2 transfers values from items to users. In Step 3, we consider the personal vocabulary again and distribute the values to items, which generates final score for each Item. Finally, we sort these scores in a descending order, and the top items having not been collected by $U_i$ will be the recommended to $U_i$.

In CTSes, different individuals have different sizes of vocabulary, and each tag may take different significance. Some tags are frequently used while some others are seldom picked. Those frequently used tags should be of higher importance in the user's viewpoint. If the user applies those frequently used tags to a specific item, it would indicate that this user prefers it to some other items assigned with infrequently used tags. Similar phenomenon also widely exists in our daily life, one can imagine that people are willing to illustrate a question using their familiar words. In addition, the number of tags assigned to an item represents how willing the user likes to describe it. By aggregating the fractions of all the tags labeling a specific item, one can estimate the importance of this item.

### 3.2 Data Set

We use a benchmark dataset, *Del.icio.us*, to evaluate the proposed algorithm. *Del.icio.us* is one of the most popular social bookmarking web sites, which allows users not only to store and organize personal bookmarks (URLs), but also to look into other users' collections and find what they might be interested in by simply keeping track of the pools with same tags or items. The data used in this paper is crawled from the website http://del.icio.us/ in May 2008. We guarantee that each user has collected at least one item, each item has been collected by at least two users, and assigned by at least one tag. Table 1 summarizes the basic information of the data set.

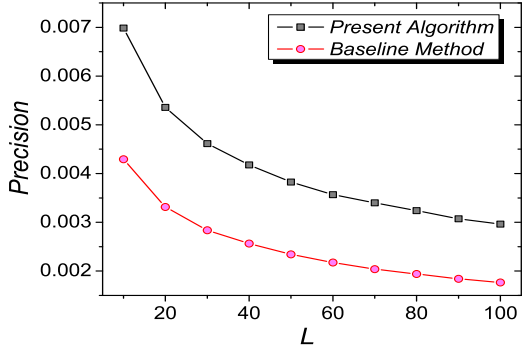

Figure 1: *Precision* versus the length of recommendation list. The results reported here are averaged over 10 independent runs, each of which corresponds to a random division of training set and testing set.

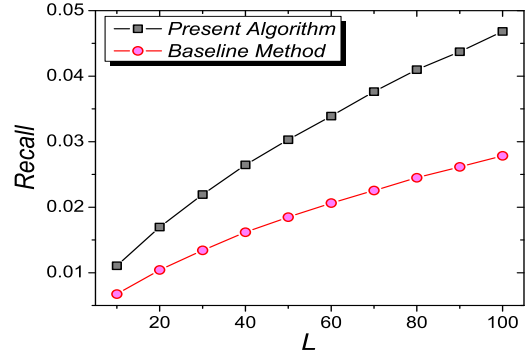

Figure 2: *Recall* versus the length of recommendation list. The results reported here are averaged over 10 independent runs, each of which corresponds to a random division of training set and testing set.

### 3.3 Experimental Results

To test the algorithmic performance, the data set is randomly divided into two parts: the training set, which is used as known information, contains 95% of entries, and the remaining 5% of entries constitute the testing set. We employ three metrics to characterize the algorithmic accuracy: *Precision*, *Recall* and *F1*, which are defined as follows [8]:

$$Precision = \frac{\sum_i N_r^i}{nL}, \qquad (4)$$

where $n$ is the number of users, $L$ is the length of recommendation list, and $N_r^i$ is the number of recovered items in the recommendations for user $U_i$.

$$Recall = \frac{\sum_i N_r^i}{\sum_i N_p^i}, \qquad (5)$$

where $N_p^i$ is the number of items collected by user $U_i$ in the testing set.

$$F1 = \frac{2 * Precision * Recall}{Precision + Recall} \qquad (6)$$

Figure 1, Figure 2 and Figure 3 show the experimental results of *Precision*, *Recall* and *F1* respectively. Since the typical length for recommendation list is tens, our experimental study focuses on the interval $L \in [10, 100]$. For comparison, we choose the method described in [25] as the baseline algorithm. It can be seen that our proposed algorithm considering the personal vocabulary significantly outperforms the baseline method in all the three measurements.

### 4. CONCLUSION AND DISCUSSION

In this paper, we proposed an tag-based algorithm that takes into account the personal vocabulary. Our algorithm is based on the following hypotheses: (i) Tags assigned to a certain item by a particular user represent personal tastes of it. Even for the same item, different individuals may give different tags. (ii) Different tags plays different roles for the same user. The frequency of tags might suggest the personal preferences: the higher the frequency, the more the user likes it. Experimental results demonstrate that the usage of tag information can significantly improve accuracy of personalized recommendations.

Recently, the collaborative tagging systems have attracted more and more attention both in the scientific and engineering worlds [4, 23]. A great number of publications and web applications have discussed/adopted tagging functions. Our experimental results show that tags can be used to not only assist personal resources organizing, but also help to filter out mass information. This paper only provides a simple way to consider the use of tags, and a couple of open issues remain for future study. From the perspective of human dynamics, the rank of tags within a single collection and the time the user chooses tags could also be taken into account. In addition, the *hypergraph* [6] description is a promising tool to exploit a comprehensive view of CTSes and bring us an in-depth understanding to the structure and evaluation of CTSes.

### 5. ACKNOWLEDGEMENT

We acknowledge Miss. Linyuan Lü for the helpful discussion. This work is partially supported by the Swiss National Science Foundation (Project 200020-121848).

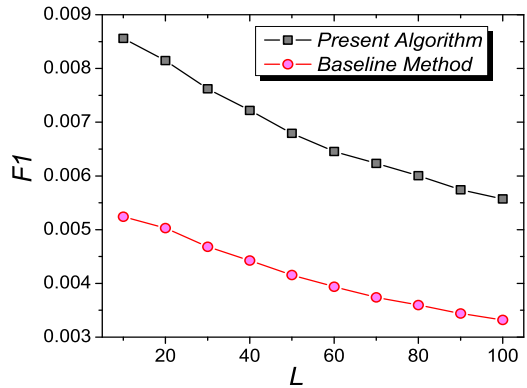

**Figure 3:** *F1* versus the length of recommendation list. The results reported here are averaged over 10 independent runs, each of which corresponds to a random division of training set and testing set.